# Miniaturized Computational Photonic Molecule Spectrometer

Yujia Zhang[1], Xuhan Guo[1], Tom Albrow-Owen[3], Zhenyu Zhao[1], Yaotian Zhao[1], Tawfique Hasan[3], Zongyin Yang[2] and Yikai Su[1]

1. State Key Laboratory of Advanced Optical Communication Systems and Networks, Department of Electronic Engineering, Shanghai Jiao Tong University, China
2. College of Information Science and Electronic Engineering, Zhejiang University; Hangzhou, China.
3. Department of Engineering, University of Cambridge; Cambridge, CB3 0FA, UK.

E-mail: guoxuhan@sjtu.edu.cn;

Miniaturized spectrometry system is playing an essential role for materials analysis in the development of *in-situ* or portable sensing platforms across research and industry. However, there unavoidably exists trade-offs between the resolution and operation bandwidth as the device scale down. Here, we report an extreme miniaturized computational photonic molecule (PM) spectrometer utilizing the diverse spectral characteristics and mode-hybridization effect of split eigenfrequencies and super-modes, which effectively eliminates the inherent periodicity and expands operation bandwidth with ultra-high spectral resolution. These results of dynamic control of the frequency, amplitude, and phase of photons in the photonic multi-atomic systems, pave the way to the development of benchtop sensing platforms for applications previously unfeasible due to resolution-bandwidth-footprint limitations, such as in gas sensing or nanoscale biomedical sensing.

**Introduction**

Spectroscopy as one of the most profound characterization techniques is attracting significant attention in astronomy, chemistry, biology, material analysis, medicine, and pharmacy, wherever light interacts with matter[1-4]. In recent years, the pursuit of high integration density and low cost become a high priority, leading to the emergence of a variety of miniaturized spectrometers that rely on the dispersion of incident light through dispersive optics such as gratings[5-7], photonic crystals[8,9], and metasurface[10,11], or through narrowband filters (array) [12-17]. However, the small fabrication tolerance and physical-size-constraint resolution severely limit the spectrometer application, especially when the device footprint is scaled down to sub-millimeter.[18]

Computational spectrometers have burgeoned thereafter to circumvent these restrictions. Leveraging more readily available computer processing power, advanced algorithmic techniques even machine learning are incorporated to decode input spectral information based on the response matrix from strong light-matter interaction, where the response matrix predominantly determines performances of computational spectrometers such as resolution, bandwidth, and reconstruction accuracy. To perfectly reconstruct the spectrum, there are some general criteria for the response matrix that need to be satisfied simultaneously: *(1) Each transmission in the response matrix should be sufficiently sharp. This sharpness enables the spectrometer to detect minor spectral features of the input signals, contributing to the high reconstruction resolution. (2) Each transmission should exhibit disorder and randomness with diverse spectral features. In other words, periodicity should be negligible across operation*

*bandwidth. In the case of transmission possessing strong periodicity, we consider any two wavelengths $\lambda_j = \lambda_i + m \cdot P$, where P refers to periodicity. For any transmission $T_p$ and $T_l$ at wavelengths $\lambda_p$ and $\lambda_l$ that satisfy $\lambda_p - \lambda_j = \lambda_l - \lambda_i$, $T_p \approx T_l$ is resulted, which means information in these points is unidentifiable in reconstruction. (3) Each transmission should be independent to each other. Reconstruction capability is determined by the orthogonality of the response matrix.* One of the major classes of computational spectrometers focuses on spectral-to-spatial mapping through dispersive components such as disorder photonic chip[19-22], stratified waveguide filters[23], and multimode spiral waveguides[24,25]. As indicated by design criterion (1) and (3), a large footprint is inevitable to accumulate sufficient optical path length (OPL) difference to provide satisfactory resolution and orthogonality. For further miniaturization of the spectrometer, resonators such as microrings become promising candidates because the narrow and sharp resonance peaks with high Q-factors are achievable in compact footprints to deliver high resolution. Nevertheless, the inherent free-spectral-range (FSR) restricts the operating bandwidth to a narrow wavelength range due to the periodicity, as demonstrated in design critirion (2). One effective approach is to incorporate double transmission modes with slightly different periodicity in order to mitigate overall periodicity, which can be achived via higher-order modes[26] or dispersive couplings[27]. Relatively high resolutions measured in dozens of picometers are realized in a wide operation bandwidth. However, the occurrence of periodicity, when not completely mitigated, may place a heavy burden on the computation cost and accuracy. Furthermore, the pursuit of high Q-factors for high resolution is often accompanied by small fabrication tolerance.

Here we propose a computational spectrometer based on a heteronuclear tetratomic PM to simultaneously satisfy design criteria (1) to (3). This PM system consists of four microdisk photonic atoms (PA) with varying radii and coupling gaps between atoms and bus waveguides. The interference between the numerous atom orbitals (AO) supported by each PA results in the formation of photonic molecular orbitals (MO) with a dispersive splitting strength that is dependent on the wavelength. Leveraging intermixing of MOs that possess diverse spectral characteristics, the mode-hybridization effect induced super-modes experiencing reinforced wavelength dependence. Thus, we successfully generated a response matrix encompassing diverse dense peaks and exhibiting disorder behaviors with almost completely eliminated periodicity. Furthermore, based on the significantly reduced scattering loss of microdisks and further enhanced transmission, such as the presence of Fano resonance with asymmetrical shape, the sharpness of the response matrix is guaranteed to deliver high resolution with a large fabrication tolerance. We experimentally demonstrate an ultra-high spectral resolution of 8 pm within a wide operation bandwidth of 100 nm. A record high bandwidth-to-resolution ratio of 1.25×10$^4$ and a record low resolution-footprint-product of 28 nm·μm$^2$ are obtained in an ultra-compact footprint measuring 70×50 μm$^2$ with merely a single spatial channel.

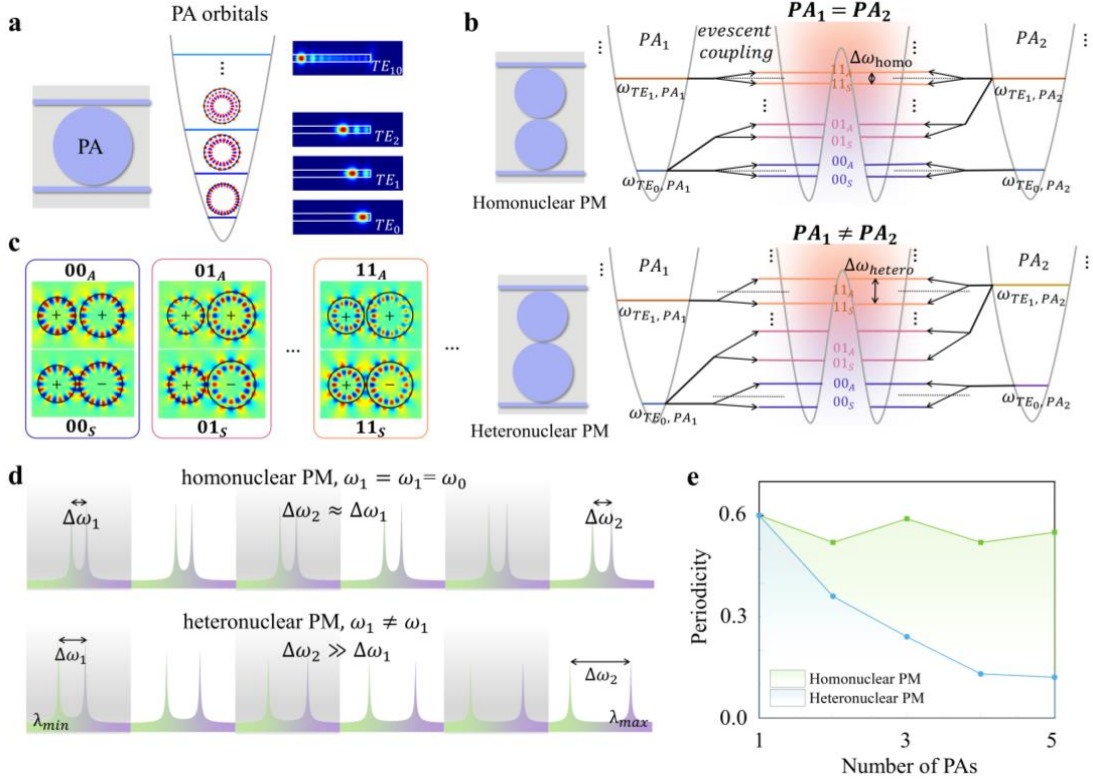

Figure 1. (a) Microdisk PA and supporting WGMs of $TE_0$, $TE_1$, $TE_2$, and $TE_{10}$, corresponding to photonic AOs with different energy levels. (b) Photonic MOs formation due to splitting mechanism of homo- and hetero-nuclear diatomic PMs. (c) Magnetic field profiles of super-modes of $SM_{00S}$, $SM_{00A}$, $SM_{01S}$, $SM_{01A}$, $SM_{11S}$, and $SM_{11A}$, where SM denotes super-modes, A/S denotes symmetric and anti-symmetric, and the number denotes the order of modes of PAs. (d) Spectral responses of splitting over wavelength of homo- and hetero-nuclear PM. (e) Periodicity as a function of the number of PAs in homo- and hetero-nuclear PMs.

**Results**

Optical micro-resonators provide promising prospects for the generation, examination, and application of confined photon states, similar to the confined electron states observed in atoms[28]. Due to this similarity, we identify the individual microdisk cavity as a photonic atom. The microdisk is a traveling wave-integrated resonator based on light total internal reflection along the boundary, which supports numerous whispering-gallery-modes (WGM) purely confined by the outer medium interface with rotational symmetry. It's quite institutively to recognize the analogy between energy levels of atomic orbitals (AO) and resonant WGMs. Figure 1(a) sketches a microdisk atom with a radius of 10 μm supporting multiple WGMs, corresponding to different AOs. The transverse-electrical (TE) polarization of WGMs from $TE_0$ to $TE_2$, and $TE_{10}$ are also provided. These WGMs possess different propagation constants, hence spectral features such as resonance wavelength ($\lambda_{res} = n_{eff}L_{rt}/m, m = 1,2,3,...$), full width at half maximum(FWHM, $FWHM = \lambda_{res}/Q$) and free-spectral range(FSR, $FSR = \lambda_{res}^2/n_g L_{rt}$), are wavelength dependent, contributing to the discriminability of atomic energy levels in PA.

With reference to the notion of PA, we then extend this analogy to photonic molecules (PM). The PM originates in the electromagnetic couplings between PAs that are arranged in close proximity. Figure 1(b) illustrates the optical MO formation of heteronuclear PM with two distinct PAs, with a homonuclear PM consisting of two identical PAs for comparison. When interfering with adjacent PAs through evanescent field couplings, mode degeneracy is consequently lifted and individual WGMs split into blue-shifted anti-symmetry super-modes ($SM_{AS}$) and red-shifted symmetry super-modes ($SM_S$) due to the constructive (destructive) interference, which is analogous to the bonding and anti-bonding MOs in diatomic molecules. Individual WGMs can interfere with other WGMs of the same order or different orders to split into MOs with a number that is larger than AOs. For example, $TE_0$ in PA1 can interfere with $TE_0$, $TE_1$, or even higher-order WGMs in PA2. Considering homo- and hetero-nuclear PM that two PAs in heteronuclear PM support AOs with different energy levels ($\omega_{TE_n, PA_1} \neq \omega_{TE_n, PA_2}$) because of different optical propagation constants, while PAs in homonuclear PM supports identical AOs ($\omega_{TE_n, PA_1} = \omega_{TE_n, PA_2}$). The splitting resulting from the interaction between $AO_{TE_m, PA1}$ and $AO_{TE_n, PA2}$ in heteronuclear PM is distinct from the splitting from $AO_{TE_n, PA1}$ and $AO_{TE_m, PA2}$, where m≠n. However, in the case of homonuclear PM, the splitting in these two conditions remains the same. In other words, mode degeneracy is not fully lifted in homonuclear PMs. The inflation in the number of MOs in heteronuclear PMs can enhance the density of discrepant resonance peaks, consequently leading to an improvement in the overall spectral sharpness and diversity. The generated magnetic field profile of super-modes of $SM_{00S}$, $SM_{00A}$, $SM_{01S}$, $SM_{01A}$, $SM_{11S}$, and $SM_{11A}$ in heteronuclear PM are depicted in Figure 1(c).

We then investigate the wavelength dependence of the splitting process for homo- and hetero-nuclear PM. Similar to diatomic molecules, the energy level difference between the symmetry and anti-symmetry MO is primarily dominated by the overlapping of the atoms, which is the inter-cavity coupling strength $k_c$ in PMs as:

$$E_\pm = \frac{\beta_1 + \beta_2}{2} \pm \sqrt{\left(\frac{\beta_1 - \beta_2}{2}\right)^2 + k_c^2} \qquad (1)$$

where $\beta_n = \omega_n - i\gamma_n$ illustrates the initial resonant frequency $\omega_n$, and $\gamma_n$ is the total optical loss of the $n$th cavity[29]. $E_\pm$ refers to the eigenvalues, whereas the real and imaginary parts indicate the super-modes' resonant wavelength and bandwidth, respectively, while ± refers to the symmetry and anti-symmetry modes. Considering the homonuclear PM possess $\beta_1 = \beta_2 = \beta_0 = \omega_0 - i\gamma_0$, splitting strengths are linearly proportional to the mutual coupling as $\Delta\omega_{homo} = \omega_+ - \omega_- = 2k_c$ and remain relatively constant regardless of wavelength change. The corresponding spectral response of homonuclear PM from drop port is exhibited in the first row of Figure 1(d). Over a large wavelength range, the splitting strength difference is negligible ($\Delta\omega_{homo,2} \approx \Delta\omega_{homo,1}$).

For heteronuclear PM, resonance wavelengths of WGMs in different cavities exhibit distinct values. The eqn.(1) can be re-expressed as:

$$E_\pm = \omega_{ave} + i\gamma_{ave} \pm \sqrt{(\omega_{dif} + i\gamma_{dif})^2 + k_c^2} \qquad (2)$$

where $\omega_{ave} = (\omega_1 + \omega_2)/2$, $\omega_{dif} = |\omega_1 - \omega_2|/2$, $\gamma_{ave} = (\gamma_1 + \gamma_2)/2$, and $\gamma_{dif} = |\gamma_1 - \gamma_2|/2$. Assuming equivalent optical loss coefficients ($\gamma_1 = \gamma_2$) in two cavities, for the resonance wavelength, the splitting strength can be estimated as:

$$\Delta\omega_{hetero} = 2\sqrt{\left(\frac{\omega_1 - \omega_2}{2}\right)^2 + k_c^2} \qquad (3)$$

which is determined by the resonant wavelengths difference of two WGMs and $k_c$

simultaneously. As demonstrated in the second row of Figure 1(d), the splitting depth to be strengthened significantly over wavelength ($\Delta\omega_{hetero,2} \gg \Delta\omega_{hetero,1}$) which is forced by the resonance wavelength difference ($\omega_1 \neq \omega_2$). This wavelength dependence in heteronuclear PM is crucial to de-periodization for discriminating each wavelength channel across a large operation bandwidth that far exceeds a single FSR.

In heteronuclear PM, the wavelength dependence is not the only factor that influences the overall spectral response. The inter-cavity coupling coefficient $k_c$, also exhibits significant variation during the formation of different MOs. Due to different degrees of phase match and distances of mode distributions from the edge of the microdisk, any single WGM can exhibit exclusive coupling efficiencies and splitting strengths when interfering with other WGMs, resulting in the reconstitution of MOs experiencing different patterns of wavelength dependence. In this way, a greater degree of diversity is introduced into the spectral response via intermixing of these MOs, leading to an effectively weakened periodicity (see Supplementary Information S1 for details).

In order to further enhance the complexity and diversity of spectral response while simultaneously eliminating periodicity, we further increase the number of PAs and propose the heteronuclear tetratomic PM comprised of four microdisk atoms. PAs with different radii are arranged in a random configuration, with different gaps between each other and the bus waveguides. This variation in geometry and topology breaks spatial symmetry and fully lifts mode degeneracy. To better illustrate the phenomenon of periodicity suppression as the number of distinct PAs increases, a three-dimensional finite-difference time-domain (3D-FDTD) method is utilized to simulate the spectral response of homonuclear and heteronuclear PMs comprising one to five microdisk PAs. Auto-correlation function is used to evaluate the hidden periodicity of the PMs. We use figure of merit (F.O.M) to evaluate the periodicity of spectra, which is defined by the maximum value in the normalized auto-correlation function after falling from the initial value "1". The periodicity of homo- and hetero-nuclear PM with the number of PA component increase from one to five are calculated and plotted in Figure 1(e) (see Supplementary Information S2 for transmission spectra and auto-correlation functions). It can be observed that the increase of PAs number in homonuclear PM has almost no suppression on the periodicity of the response spectrum, due to the insensitivity of wavelength during super-modes splitting process and the insufficient lifting of mode degeneracy. However, for the tetratomic or pentatomic heteronuclear PM, the augmentation of distinct microdisk PAs with random configuration in PM results in the explosive growth of optical super-modes due to mode hybridization effect, meanwhile the wavelength dependence of these super-modes is further catalyzed. Hence, periodicity is gradually suppressed to a negligible level (see Supplementary Information S3 for more characterization). In this case, a unique spectral response with high-density resonance peaks is successfully generated, exhibiting a complete disorder and diverse features. The tetratomic heteronuclear PM can effectively triumph over the inherent periodicity restriction over a large operation bandwidth, which is ten times the initial FSR of a single WGM, while simultaneously maintaining the sharp resonance features in the transmission spectrum.

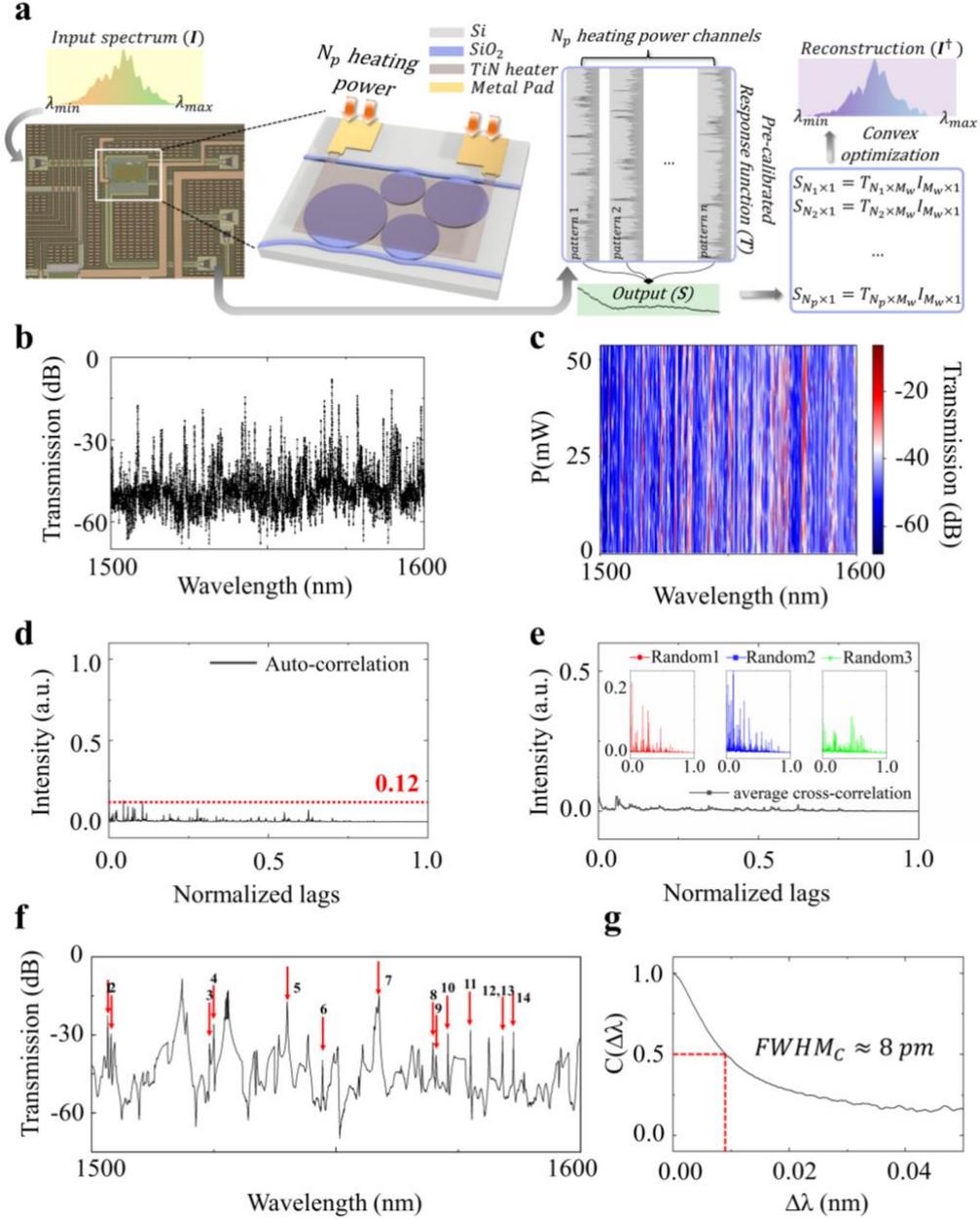

Figure 2. (a) Reconstruction process. The inset shows the optical microscopy photo of the fabricated device. (b) Transmission spectrum under zero heating powers. (c) Response matrix under linearly swept heating powers from 0 mW to 60 mW. (d) Auto-correlation function. Red dashed line labeled the periodicity level. (e) Cross-correlation function. Average cross-correlation function is plotted as black line. Insets show the cross-correlation function of three groups of transmissions that are randomly picked, labeled as red, blue and green. (f) Zoom-in transmission spectrum from 1540 nm to 1550 nm. Red arrows and numbers marked some resonance peaks. (g) Zoom-in plot of $C(\Delta\lambda)$ around $\Delta\lambda = 0$. Red dashed line depicts the estimated resolution.

## PM Spectrometer characterization

The schematic layout of the proposed spectrometer based on a heteronuclear tetratomic photonic molecule (PM) and the reconstruction process are illustrated in Figure 2(a). The four microdisks atoms are with radii of 10 μm, 14 μm, 13 μm, and 18 μm, respectively. The gaps between microdisks and bus waveguides are randomly set around 180 nm for strengthening randomness and coupling strengths diversity. The designed tetratomic PM spectrometer is fabricated in CUMEC using their standard multi-project wafer (MPW) tape-out process. It's worth noting that, the imperfection induced by fabrication such as localized defects, layer thickness variation, deposition nonuniformity, and etching roughness, will further lead to random deviation in coupling efficiency, resonant wavelength, and spectral shape. In this respect, our tetratomic PM spectrometer exhibits outstanding fabrication tolerance. The inset in Figure 2(a) shows the optical microscopy photo of the fabricated on-chip random spectrometer, where the functional microdisks region is only with a footprint of $70 \times 50\ \mu m^2$. The PM is subject to synchronous regulation through a TiN heater on top of the microdisks and coupling regions by means of the thermos-optic (TO) effect (see Supplementary Information for optical microscopy photos with and without heater).

For the computational reconstruction process, reconstruction matrix $T$ with a row number of heating power channel $N_p$ and a column number of wavelength points $M_w$ is pre-calibrated by measuring transmission spectra in the output port while linearly sweeping the external heating power. If the unknown incident signal $I$ with $M_w$ wavelength points pass through the tetratomic PM spectrometer, the spectral information can be sampled and encoded, along with the detected output power $S$ with $N_p$ heating channel number recorded. $S$ can be mathematically expressed as a vector product as:

$$S_{N_p \times 1} = T_{N_p \times M_w} I_{M_w \times 1} \tag{4}$$

By harvesting $N_p$ group of transmission response, where $N_p$ is far less than Nyquist frequency, a signal with $M_w$ wavelength points can be reconstructed. The reconstructed signal $I^\dagger$ can be located by solving this under-determined linear least-squares:

$$I^\dagger = argmin_{I \in \mathbb{R}^+} \|TI - S\|_2^2 \tag{5}$$

Where $\|\cdot\|_2$ represents $l_2$-norm.

A pre-calibration process is demonstrated as follows. A tunable continuous wave laser (Santec TSL 770) and a power monitor (Santec MPM 210) are utilized for sampling and data collecting. A source meter (Keithley 2400) is used to offer an external driving power source. A maximum external power $P_{max}$ of 60 $mW$ is applied to the heater. The dimensions of response function $T$ are determined by the wavelength point number $M_w = BW/\delta\lambda$, and $N_p = P_{max}/\delta P$ mutually, where $\delta P$ is the sampling step of 0.3 mW; BW is the spectrometer measurable bandwidth that is from 1500 nm to 1600 nm; $\delta\lambda$ is the wavelength grid of 10 pm. Here, a spectrum with $M_w$ of 10000 points can be reconstructed with merely $N_p$ of 200 heating power channels, demonstrating an ultra-high compress ratio. Figure 2(b) shows the measured response under $P = 0\ mW$, which is the initial transmission spectrum from the output port. The sampled response function under different external powers is exhibited in Figure 2(c).

In the response matrix, some coherent interference effects such as Fano-shape resonance and coupled-resonator-induced transparency (CRIT), which can be analogous to electromagnetically induced transparency (EIT) in atom system are frequently witnessed over the whole operation bandwidth (see Supplementary Information S4). The presence of these coherent effects improves diversity within the response matrix and Q-factors of these resonance peaks are effectively enhanced, resulting in improved overall spectral sharpness. Furthermore. due to different temperature sensitivity yielded by different WGMs, inter-cavity coupling $k_c$ is strongly manipulated between different channels. The splitting strength of each MO is dependent on heating power, resulting in spectral signatures that varies with power channels, which is far beyond merely resonant red shifting (see Supplementary Information S4). The implementation of this mode evolution yields an impressive benefit in terms of the orthogonality of the response matrix, which suppress cross-correlation effectively. This enables the reconstruction of incident signals that possess a large number of wavelength points within limited heating power channels.

We further use auto-correlation to evaluate the hidden periodicity level, and cross-correlation to identify the degree of independence between any two transmission spectra of the measured response matrix. The average auto-correlation has been computed and is presented in Figure 2(d). The F.O.M of auto-correlation symbolizing periodicity is below 0.12, which is indicated by a red dashed line. Clearly, the instinct periodicity of a single cavity is fully fractured and becomes imperceptible in the fabricated PM spectrometer. The calculated average cross-correlation is marked as black scatter dots, and the other three cross-correlations of arbitrarily picked three transmission spectra marked are marked as red, blue, and green scatter dots respectively in Figure 2(e). It is observed that the cross-correlation is also maintained at a relatively low level, indicating a nearly mutually orthogonal response matrix.

The calculated Q-factors and bandwidths of some resonance peaks are marked in Figure 2(f) and summarized in Table.1. As the coupling and interference effects in our tetratomic heteronuclear PM are too complicated, Q-factors for each resonance peak vary significantly. Within the spectral range spanning from 1540 nm to 1550 nm, the loaded Q value for our PM reaches a maximum of $7.74 \times 10^5$, corresponding to a bandwidth of 2 pm. Such high Q-factors are the foundation for achieving high reconstruction resolution, as illustrated in design criterion (1). There exist some reasons for such high Q-factors. Firstly, the microdisk facilitates the propagation of WGMs that are entirely confined by the surface of the outer medium because of the total internal reflection of light. It can be observed that the microdisk cavities consistently demonstrate higher Q-factors in comparison to microring resonators of similar OPL. This is primarily attributed to the significant decrease in Rayleigh scattering loss, which arises from the existence of only one rough sidewall during fabrication. Secondly, in comparison to single-mode resonators, the inclusion of multiple higher-order WGMs and the generation of numerous super-modes result in a greater number of resonance peaks distributed across the spectrum. This leads to an enhancement of fluctuations, consequently increasing the overall sharpness of response matrix, denoted as $\sum_0^{i=M_w}|T_i - T_{i+\delta\lambda}|$. Lastly, the presence of CRIT and Fano-type asymmetric resonance peaks, which exhibit a wide distribution across the spectra, serves to increase the trapping time of photons within cavities and results in the sharpening of resonance peaks.

Table.1 Bandwidth and Q-factors of some resonance peaks in the measured transmission spectrum

| Peak serial number | 1 | 2 | 3 | 4 | 5 | 6 | 7 | 8 | 9 | 10 | 11 | 12 | 13 | 14 |
|---|---|---|---|---|---|---|---|---|---|---|---|---|---|---|
| Bandwidth (pm) | 5.6 | 8.1 | 5.7 | 11.2 | 11.1 | 5.5 | 12.6 | 6.9 | 10.8 | 4.9 | 3.5 | 3.1 | 2.0 | 3.0 |
| Loaded Q ($\times 10^5$) | 2.75 | 1.90 | 2.71 | 1.38 | 1.39 | 2.81 | 1.23 | 2.24 | 1.43 | 3.16 | 4.42 | 4.99 | 7.74 | 5.16 |

For computational photonic molecule spectrometer, the estimated resolution based on the response matrices can be numerically calculated by:

$$C(\Delta\lambda, N) = \frac{\langle T(\lambda, N)T(\lambda + \Delta\lambda, N)\rangle}{\langle T(\lambda, N)\rangle\langle T(\lambda + \Delta\lambda, N)\rangle} - 1 \tag{6}$$

where $T(\lambda, N)$ refers to spectral transmission at the $N^{th}$ bias for input wavelength $\lambda$, $\Delta\lambda$ is the spectral spacing of two wavelength points, $\langle...\rangle$ refers to the average over $\lambda$. $C(\Delta\lambda, N)$ is plotted in figure 2(g). The estimated reconstruction resolution is full width at half maximum (FWHM) of $C(\Delta\lambda, N)$ which is approximately 8 pm, as marked in Figure 2(h). The calculated resolution largely conforms to the average value of Q-factors of numerous resonance peaks across the whole spectrum.

The computational PM spectrometer with complicated and diverse MOs is experimentally generated with a high resonance peak density. The large number of nonperiodic-appeared peaks enhances the efficacy of encoding for incident signals and facilitates greater diversity in response matrices, thereby permitting higher resolution in reconstruction in a broad operation bandwidth that far exceeds the inherent periodicity of cavities. The effectiveness of the proposed spectrometer in retrieving various probe signals with distinct optical features is numerically assessed and demonstrated (see Supplementary Information S7). Furthermore, a comprehensive analysis has been carried out to validate the robustness against measurement noise, and high levels of sampling and encoding efficiency in our tetratomic PM spectrometer (see Supplementary Information S5, S6).

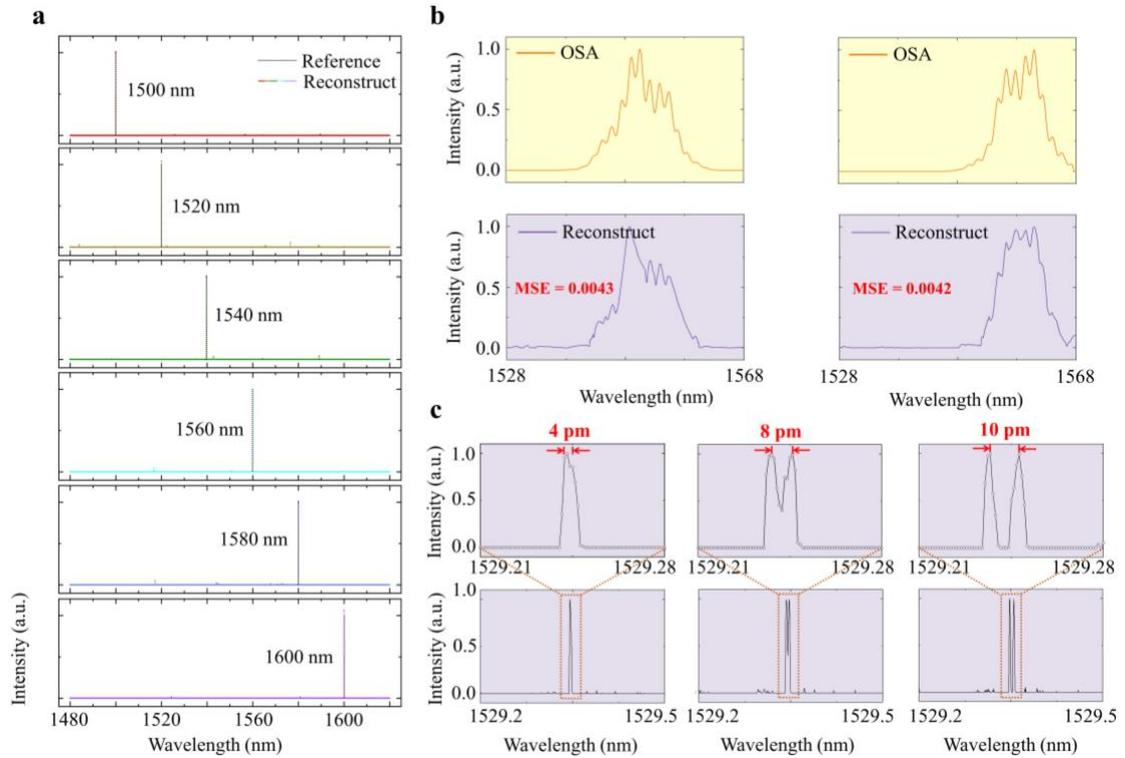

Figure 3. (a) Single narrow linewidth signal reconstructions. (b) Continuous signals reconstructions. (c) Dual narrow linewidth signal reconstructions.

For eliminating current fluctuations due to unstable contact, we utilize wire bonding and an electrical package for our fabricated chip. The current value can be stabilized at $1\times10^{-4}$ mA after electrical packaging, Broadband grating couplers (GC) with less than 5 dB insertion loss are applied for fiber-chip coupling. The pre-calibration of the response matrix is demonstrated in Figure 2(c). It is noteworthy to mention that the response matrix obtained in real-world circumstances frequently indicates a certain level of ill-conditioned, particularly in regard to the interpretation of different incident signals in the presence of measurement noise. In order to overcome this overfitting obstacle, the regularization coefficient is introduced as:

$$I^{\dagger} = argmin_{I\in\mathbb{R}^+}(\|T \times I - S\|_2^2 + \alpha_1\|I\|_1 + \alpha_1\|DI\|_2^2), 0 \leq I \leq 1 \quad (7)$$

Where $\alpha_1$ is the weight regularization coefficient for the $l_1$-norm of input matrix I, which is vital in the regression of $I^{\dagger}$ into discrete untrivial solutions; $D$ refers to the first derivative operator; $\alpha_2$ is the weight regularization coefficient for the $l_2$ norm, which is critical for continuous broad-band signals reconstruction, mainly used to smooth sharp changes and suppress measurement noises in reconstructed spectra. CVX optimization algorithm implemented in MATLAB environment is utilized to solve this convex optimization.

For single peak reconstruction, we use a tunable laser source as input signals with 1510 nm, 1530 nm, 1570 nm, and 1590 nm wavelength, respectively, throughout the whole operation bandwidth of 100 nm. The reconstruction results for these single narrow linewidth signals are presented in Figure 3(a). Signals located throughout the whole bandwidth are successfully reconstructed in accurate positions. We also characterized continuous signal reconstruction. A broadband amplified spontaneous emission (ASE) source is utilized to supply incident continuous light with a bandwidth of over 60 nm centered in 1550 nm. A programmable optical filter (Finisar Waveshaper 1000s) is connected to generate the required test signals. We use a commercial optical spectrum analyzer (OSA, Yokogawa AQ6370C) to collect optical response of the waveshaper that is coded with required signals in advance for calibration and reference, as presented in the first row in Figure 3(b). We use mean-squared-error denoted as:

$$MSE = \frac{1}{M_w}\sum_1^{M_w}(I - I^{\dagger})^2 \quad (8)$$

to evaluate the reconstruction accuracy, where $I$ is obtained from OSA. The reconstructed results are exhibited in the second row of Figure 3(b), with MSEs of 0.0043 and 0.0042. Although a few details in reconstructed spectra are twisted, the proposed multi-microdisk random spectrometer exhibits good performance when tackling discrete and continuous signals (see Supplementary Information S8 for reconstruction with higher accuracy provided by the alternative device). The primary cause of some disagreement with reference in broadband signal reconstruction stems from the spatial oscillation of multi-axis manual stages and vertical fibers. This can be effectively suppressed by the utilization of optical packaging. The utilization of temperature controllers is also considered in our future measurement for mitigating reconstruction errors caused by temperature fluctuations.

We further explore the resolution limits of our device. Here we re-calibrate and conduct reconstruction based on $\delta\lambda$ of 1 pm. We then use two tunable laser sources to generate two narrow linewidth discrete signals and then combined them with a 3-dB coupler. Figure 3(b) demonstrates the reconstruction result when these two signals are separated merely by 4 pm, 8 pm, and 10 pm. Apparently, 4 pm is beyond the resolution limit that the PM spectrometer recognized it as a single peak signal. When the two signals' wavelength distance increase to 8 pm, two peaks can be evidently distinguished, demonstrating an ultra-high resolution of 8 pm that satisfy the Rayleigh criterion. When the two signals' wavelength distance further increases to 10 pm, the discrete signals can be fully reconstructed as intact peaks. Here, we experimentally substantiate an ultra-high spectral resolution of 8 pm for the proposed multi-disk random spectrometer.

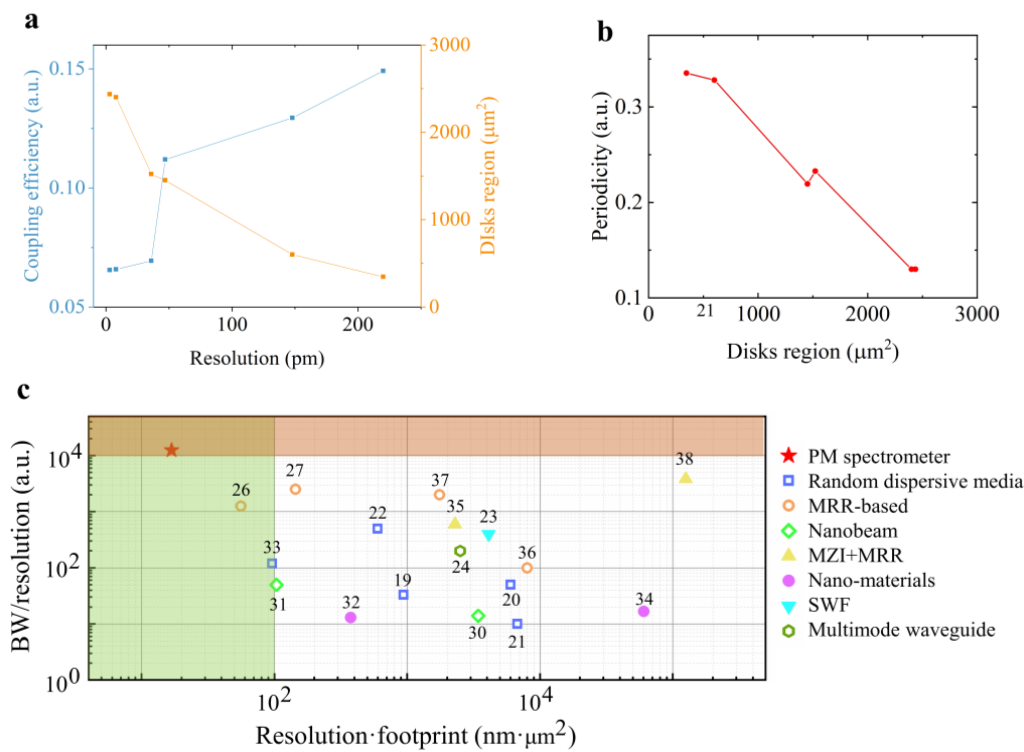

Figure 4. (a) Average coupling efficiencies and microdisk covered regions of other alternative devices as functions of estimated resolution. (b) Periodicity of other alternative devices as functions of microdisk-covered regions. (c) and (d) Performance comparison of our work and other representative computational spectrometers[19-24,26,27,30-38]. The green region indicates RFPs smaller than 100; the red region indicates bandwidth-to-resolution ratios larger than $10^4$.

The resolution of a computational spectrometer is dominated by the overall sharpness of its response function, which is reflected primarily in the Q-factors of our resonance-cavity-based device. In order to conduct a more comprehensive investigation into the deterministic impact of PM structure parameters on spectral resolution, we additionally analyzed alternative fabricated spectrometer devices that are based on heteronuclear PMs. These PMs exhibit varying coupling efficiencies between microdisks and bus waveguides, and varying microdisks regions. The reduction of average coupling strengths results in a decrease in the external losses of the resonator. Meanwhile, the expansion of microdisk regions leads to an effective increase in OPL and the number of supported AOs. As illustrated in Figure 4(a), both the reduction of average coupling strengths and the increase of microdisk regions contribute to an enhancement in the Q-factors, resulting in an optimization of the estimated resolution to 3 pm. Moreover, it can be observed from Figure 4(b) that the periodicity in PMs diminishes gradually as microdisk regions increase. This can be attributed to further quantities of super-modes generation based on the increased number of AOs in microdisks with larger radius. Consequently, this leads to an enhancement of the mode hybridization effect to break the strong periodicity inherently delivered by single or few resonant modes.

Finally, we conduct a comprehensive comparison of the primary performances of various state-of-art computational spectrometers[19-24,26,27,30-38] that have been previously reported in Figures 4(c). We emphasize the superior performance of our spectrometer in terms of resolution, operation bandwidth, and footprint. Because resolution is tightly related to footprint, the evaluation of spectrometer performance per unit footprint is conducted by defining the resolution and footprint product (RFP). The evaluation of the trade-off between resolution and bandwidth in a single spectrometer device is performed using the bandwidth-to-resolution ratio, which is one of the most representative spectrometer performance indicators. This ratio is inherently constrained by the contradiction between the larger OPL required for high resolution and the resulting decrease in bandwidth, which are mostly limited to the order of dozens or hundreds. As illustrated in Figure 4(c), the proposed PM spectrometer achieves a record-high value of the bandwidth-to-resolution ratio of $1.25 \times 10^4$ within a lowest RFP value of 28 nm·$\mu m^2$. Our PM spectrometer demonstrates an ultra-high resolution of 8 pm while maintaining a low RFP (See Supplementary Information S9 for more comparison details). Considering the operation bandwidth of 100 nm is mainly limited by the operation bandwidth range of vertical grating couplers, well-designed fiber-to-chip coupling methods or segment operation can further improve our bandwidth-to-resolution ratio. Meanwhile, wider operating bands not limited to the C-band can also be realized via a single PM spectrometer.

**Outlook**

We have demonstrated the miniaturized photonic molecule spectrometer that provides a new paradigm to achieve highly efficient spectroscopy. Enhanced by a computational spectral reconstruction algorithm, the device shows a record high bandwidth-to-resolution ratio of $1.25 \times 10^4$ and a record low resolution-footprint-product of 28 nm·$\mu m^2$, the footprint shrinks to $70 \times 50$ $\mu m^2$ with merely a single spatial channel. The computational photonic molecule spectrometer method essentially breaks the constraints of resolution and bandwidth of the conventional resonant filter array systems while maintaining the miniaturized footprint, could promise a great breakthrough to boost the chip-wise spectroscopy, which is particularly useful for *in-situ* and portable applications with high resolution such as gas sensing or nanoscale biomedical sensing.

**Data availability**

The data that is relevant to this work is available from corresponding authors on reasonable request.

**Code availability**

The code that is relevant to this work is available from corresponding authors on reasonable request.

**Acknowledgements**

This work was financially supported by the National Key R&D Program of China (2021YFB2801903); Natural Science Foundation of China (NSFC) (62175151 and 61835008). We also thank the Center for Advanced Electronic Materials and Devices (AEMD) of Shanghai Jiao Tong University (SJTU) and United Microelectronics Center (CUMEC) for the support in device fabrication.